**Towards automatic detection and classification of orca (*Orcinus orca*) calls using cross-correlation methods**

Stefano Palmero[1,*], Carlo Guidi[1,2], Vladimir Kulikovskiy[2], Matteo Sanguineti[1,2], Michele Manghi[3], Matteo Sommer[4] and Gaia Pesce[5]

[1] University of Genova, Via Dodecaneso 33, Genova, I-16146, Italy

[2] INFN, Sezione di Genova, Via Dodecaneso 33, Genova, I-16146, Italy

[3] NAUTA scientific, Strada della Carità 8, Milano, 20135, Italy

[4] Costa EDUTAINMENT S.p.A., Area Ponte Spinola, Genova, 16128, Italy

[5] Acquario di Genova, Area Ponte Spinola, Genova, 16128, Italy

[*]Corresponding author

*E-mail address*: st.palmero@gmail.com

**Abstract**

Orca (*Orcinus orca*) is known for complex vocalisation. Their social structure consists of pods and clans sharing unique dialects due to geographic isolation. Sound type repertoires are fundamental for monitoring orca populations and are typically created visually and aurally. An orca pod occurring in the Ligurian Sea (Pelagos Sanctuary) in December 2019 provided a unique occasion for long-term recordings. The numerous data collected with the bottom recorder were analysed with a traditional human-driven inspection to create a repertoire of this pod and to compare it to catalogues from different orca populations (Icelandic and Antarctic) investigating its origins. Automatic signal detection and cross-correlation methods (R package warbleR) were used for the first time in orca studies. We found the Pearson cross-correlation method to be efficient for most pairwise calculations (> 85%) but with false positives. One sound type from our repertoire presented a high positive match (range 0.62–0.67) with one from the Icelandic catalogue, which was confirmed visually and aurally. Our first attempt to automatically classify orca sound types presented limitations due to background noise and sound complexity of orca communication. We show cross-correlation methods can be a powerful tool for sound type classification in combination with conventional methods.

**Introduction**

Orcas (*Orcinus orca*) exhibit two main ecotypes mainly based on vocal activity and ecology (Ford, 1989), i.e. the resident orcas, which are mostly fish-eating and often communicate during activities, and transient orcas, which is mostly mammal-eating and less vocal than the former, since mammal prey species can detect their vocalisations (Deecke et al., 2005).

Like other marine mammals, orcas produce a variety of sounds that can be divided into three discrete types: clicks, whistles and burst-pulse calls. Clicks are very short (< 1 ms) sounds ranging from 10 to 100 kHz that are repeatedly emitted for echolocation, i.e. to navigate and hunt (Au et al., 2004). Whistles are constantly frequency-modulated tonal sounds with a bottom fundamental frequency followed by repeated harmonic overtones. These sounds can last several seconds and differ in range depending on geographic areas, e.g. 1–74 kHz in the Northern Atlantic Ocean (Samarra et al., 2010). Burst-pulse calls (hereafter calls) are rapidly repeated sounds that can contain a low-frequency component (LFC), i.e. monophonic call, or a combination of low and high-frequency components (HFC), i.e. bi-phonic call, and can be formed by more than one distinct part. LFCs are spaced according to a certain sideband interval (SBI). Calls are more intense between 0.5 and 25 kHz and generally last 0.5–1.5 seconds (Filatova et al., 2015). Both whistles and calls are believed to serve for social communication and inter-group identification (Ford, 1989; Thomsen et al., 2002).

Calls are the most common type of sound emitted by orcas thus bioacoustic research mainly focuses on this type of sound (Ford, 1991). Based on multiple studies, different orca populations present a unique repertoire of sounds resulting in dialects that are passed on throughout matrilines as a learned behaviour (Yurk et al., 2002). A group of orcas, i.e. pod, is defined by its repertoire. Pods sharing dialects form a bigger clan, typically united during foraging, and show genetic similarities (Barrett-Lennard, 2000). Therefore, the study of dialects through call classification can serve as an efficient tool for monitoring the status of different orca populations distributed worldwide.

Classification of orca sounds has been widely achieved visually and aurally (Yurk et al., 2002). To make this approach consistent, automatic classification is desired for handling larger amounts of data ensuring reproducibility. However, factors such as environmental noise can pose limitations to automatic classification, potentially resulting in biased results. As such, the method is still under development (Brumm et al., 2017). Neural networks and deep learning have already been used to automatically detect orca sounds with the purpose of species identification (Poupard et al., 2019a.) and call classification (Bergler et al., 2019; Poupard et al., 2019b). Recent studies used dynamic type warping (DTW), a method for measuring pairwise acoustic dissimilarity based on dominant or fundamental frequency contours, to replace human classification and classified orca sounds in both captivity and natural conditions (Brown et al., 2006; Brown & Miller, 2007). Cross-correlation represents another method to measure pairwise acoustic dissimilarity (Clark et al., 2010) and consists of a step-by-step comparison that is performed by sliding one spectrogram over the other thus calculating an amplitude correlation value at each step. Unlike DTW, this method is robust to different background noise levels (Cortopassi & Bradbury, 2000) is recommended when recordings have remarkable differences concerning background noise (Araya-Salas et al., 2017). The recently developed R package "warbleR" (Araya-Salas & Smith-Vidaurre, 2016) is a powerful and flexible tool allowing for the use of different methods to perform bioacoustic analysis.

Here we attempt to analyse a sample of different sound types of an orca pod occurring in the Ligurian Sea (Pelagos Sanctuary) in December 2019 using cross-correlation methods. The pod consisted of a total of four grown-up individuals recognised through photo identification (Mrusczok et al., 2021) and one calf. In particular, Riptide (SN113), an adult male, Aquamarin (SN116), a juvenile, Dropi (SN115), whose sex was not identified and Zena (SN114), an adult female who lost her calf during data collection. Individuals from this pod were regularly detected in Iceland starting from 2013 until 2018. When observed during this period, the individuals of this pod displayed foraging techniques such as carousel and subsurface feeding as well as sharking including both fish and mammal species in the diet (Mrusczok et al., 2021). The goal of this work is twofold. From one side, we used the conventional visual and aural spectrogram inspection to classify sound types and provide a repertoire of this pod as well as compare it with already existing catalogues aiming to investigate dialect sharing between pods and clans and provide acoustic evidence of the pod origin. On the other side, we use warbleR for the first time aiming to repeat the same procedure using non-human driven algorithms. In particular, we automatically identify samples in the recordings to classify sound types through cross-correlation and compare samples across catalogues.

The pod of this study was almost always stationary during its presence in the Ligurian Sea, which facilitated the data collection. Thanks to the collaboration with the Port Authority and Acquario di Genova we performed several boat campaigns supported by Coast Guard in which orca vocalisations were collected with two hydrophones. Contemporary to this, a large amount of data was collected using a wideband autonomous underwater acoustic recorder anchored to the sea bottom for several days. The data quality from the bottom recorder was much superior compared to the data collected in the sea campaigns due to the boat and sea surface noises absence, so only data from the former was included in the analysis.

We use this special opportunity to study an isolated Icelandic pod far from its area of origin thus potentially providing information on dialect development due to geographic separation. From the perspective of the data analysis, this case study aims to test whether the cross-correlation method is robust and provide parameters for facilitating its application to further datasets from other contexts.

**Material and methods**

Data used in the analysis were collected using a uRec384k programmable underwater acoustic recorder (NAUTA Scientific, Italy), belonging to a class of instruments also known as bottom recorders. This particular recorder, offering 16-bit resolution, is fit, as a standard, with a Sensor Technology SQ26-05 hydrophone whose frequency range is known to greatly outperform its specifications. While the units are described as having a sensitivity of -169 dB re 1V/uPa up to 50 kHz, they keep a reasonable sensitivity curve up to 100 kHz. For this reason, the recorder was set to sample at 192 kHz, continuously, organising the files in five-minute chunks.

The hydrophone was set from 11:29 on 07/12/2019 to 11:12 on 11/12/2019 resulting in 1325 five-minute recordings sampled at 192 kHz of which 217 (16.38 %) contained orca sounds. These 217 recordings were further divided into four categories based on the number of visually found emissions in the spectrograms as follows: (I) 1-5 killer whale sounds; (II) 6-15; (III) 16-30; (IV) >30. All recordings in the latter category with 2.30 hours total duration were visually and aurally inspected in a preliminary analysis aiming to identify all samples. In total, 1460 samples were identified and subsequently grouped into 33 sound types including whistles, calls and composite sounds (i.e. a call composed of two or more distinct parts). The bi-phonic calls "G1.1" (16%), "G1.2" (3%), "G1.3" (24%) the monophonic call "G3" (8%), and the most detected composite sound, i.e. "G2" (10%) are shown in Fig. 1. These latter sounds were selected for the detailed cross-correlation study thanks to the abundant presence of their clean samples in the data. We used 16 five-minute recordings in (II-IV) categories with the lowest background noise matrix to manually create a dataset of sample locations (start/end time) within the recordings considering the five above-mentioned sound types.

This information was used as an input for the warbleR function "auto_detec", which automatically finds the given samples within recordings. For the automatic approach, we also tested the auto_detec function on three five-minute recordings but without information on the sample locations. For both approaches, only the samples with an amplitude threshold of 5% and a minimum length of 0.2 seconds, i.e. long signals (Araya-Salas et al. 2017) were included to avoid misidentifications in line with results of the "optimize_auto_detec" function performed over several tests. According to Araya-Salas and Smith-Vidaurre (Araya-Salas et al., 2017), samples with a signal to noise ratio (SNR) < 1 should be discarded from the analysis as the signal cannot be distinguished from the background noise matrix. For this analysis, we used an even more stringent SNR threshold, i.e. SNR ≥ 5, to further reduce this bias.

Concerning the manual approach, cross-correlation was performed on the best five samples, i.e. with the highest SNR, of each sound type. We used the warbleR "cross_correlation" function with the Pearson correlation coefficient (PCC) for pairwise calculations by computing a fast Fourier transform (FFT) in a rectangular Hann window of length 1024 with a bandpass filter of 0.2–9 kHz, i.e. cutting off bottom noise and yet maintaining relevant spectrogram features including LFCs and HFCs for comparisons, and a 90% overlap for more accurate results (Araya-Salas & Smith-Vidaurre, 2016) (see Fig. 1 for examples of spectrograms). Additionally, we modified the cross_correlation function implementing FFT of each of the signals from the manual approach and pairwise $chi^2$ calculation in the same time and frequency range as for the Pearson cross-correlation and compared results across methods. Based on the correlation matrix resulting from the manual approach, i.e. active matrix, we used the "mmds" function of the "bios2mds" package (Pelé et al., 2012) to calculate Principal Coordinate Analysis (PCoA) (Gower 1966) using *1 - PCC* and *$chi^2$ - 1* as a distance metric for the respective methods. Each sample from the automatic approach was subsequently related to the active matrix as well as other new samples through a one-vs-all cross-correlation with the same parameters mentioned above followed by the calculation of PCoA.

As a final step, we selected eight samples including both monophonic and bi-phonic calls, i.e. "I40.1", "I42.1", "I47", "I53.1", "I53.2", "I62.1", "I64" and "I71", and one composite sound, i.e. "I43.1", of the Icelandic catalogue (Selbman et al., 2019) being visually and aurally compatible to those from our repertoire and having the same sampling rate, i.e. 192 kHz, allowing to perform cross-correlations. In this case, we only compared samples performing a Pearson cross-correlation. Visualisation of PCoA in space is not presented here for simplicity. Following Araya-Salas and Smith-Vidaurre (Araya-Salas & Smith-Vidaurre, 2016) and based on the results of several tests, we set a correlation coefficient of 0.50 and 0.60 for low (≥ 0.50) and high (≥ 0.60) positive matches, respectively. Accordingly, values < 0.50 indicated two samples could be considered as belonging to different sound types.

Apart from the analysis in the R software, we visually compared our repertoire with the above-mentioned Icelandic catalogue as well as one from Antarctic orcas (Wellard et al., 2020), expecting clear differences concerning the latter due to remarkable geographic separation.

*Use of experimental animals*

No animals were used or handled in the study.

**Results**

*Visual analysis*

Here we report the visually and aurally created repertoire of our orca pod with a total of 23 sound types including whistles, i.e. type G9, bi-phonic calls, i.e. type G1, monophonic calls, i.e. types G3, G4, G5, G6, G7 and G8 and composite sounds, i.e. types G1, G2 and G9, with the percentage of occurrence (Fig. 1 includes results of the automatic approach which are described below). Furthermore, we compare our repertoire to those from the Icelandic and Antarctic catalogues.

G1.1 and G1.2 can be considered a likely match as they look very similar concerning all spectrogram features, except for the very first section of HFCs while G1.3, G2, G3, G5.4, G6 and G8 are distinct sound types. G1.3 presents a variation with LFC starting to oscillate over time (G1.5). G4.1 and G4.2 only differ concerning the SBI, same for G7.1 and G7.2. The first part of the composite sounds matches with G1.3 but the former is also accompanied with a whistle of type G9 (G1.4), very dense clicks (G1.7) or both occurring in differing order (G1.8 and G1.9). G1.6 is a unique mix of 1.3, clicks and G1.5.

We noted that G1.2 and G2 separated by about one second occurred as a pattern (Supp. Fig. 1) 140 times in the analysed recordings. Similarly, G1.3 followed by G1.1 occurred 35 times and G4.1/G4.2 followed by G7.1/G7.2 occurred 33 times.

Concerning Icelandic samples, the monophonic calls I40.1 and I43.1 are chosen as visually compatible with the first part of the composite sound G2 although the SBI differs from that of the latter. I42.1 presents similar LFCs and HFCs to G1 and in particular, the SBI is closer to G1.3. The bi-phonic call I47 is visually similar to G5, especially with G5.3 and G5.4 concerning the SBI and LFCs. I53.1 visually and aurally matches with G4.2. I53.2 visually and aurally matches with G1.3 although their HFCs follow different trends. I62.1 is visually and aurally similar to G1.3 but with slightly different SBIs. I64 is visually and aurally similar to G4.1 although the sample of the latter is shorter, i.e. about 0.25 and 1 second for G4.1 and I64, respectively. I71 and G3 are similar concerning the overall decreasing trend of LFCs although SBIs are different.

In total, we noticed that about 19 sound types out of 74 in the Icelandic catalogue can potentially match sound types from our catalogue. In particular, I53.1 and I53.2 can be considered as likely matches with G4.2 and G1.3, respectively. For the remaining sound types, the differences in the SBI and trend of both LFCs and HFCs show the signals do not match exactly.

Concerning the Antarctic catalogue, we found six over a total of 28 sound types are comparable to our samples. In particular, part of the composite sound McM1 with its subcategories as well as McM6 are similar to type G5, although the SBI is different in the Antarctic samples. The composite sound McM3 is partly similar to G1.3 including SBIs although the trend of HFCs is different. Even with a different general shape, the composite

McM5 with its subcategories and McM9 are comparable to G3. The whistle McM7 is similar to G1.6 with, however, a different SBI. Therefore, from the visual analysis, there is no evidence of matches between our samples and those from the Antarctic catalogue.

*Cross-correlation analysis*

Concerning the manual approach, not all the samples in the dataset were detected by *auto_detec* because of poor quality, resulting in a smaller effective number of samples per group (Table 1).

The selection was further reduced when keeping only samples with SNR ≥ 5. With the second approach, we found a total of 71 samples in the three recordings. Of these, only 46 had SNR ≥ 5. In particular, 16 samples over 46 were previously found with the manual approach as belonging to the five specific sound types ("G1.1", "G1.2", "G1.3", "G3", "G2"). A total of 33 samples of the same sound types were manually identified meaning only about 50% of the samples were automatically re-found with the automatic approach.

Inter- and intra-group correlation coefficients resulting from the Pearson cross-correlation of all samples of the manual approach with SNR ≥ 5 are shown in Table 2 and Supp. Fig. 2. In particular, applying a threshold of ≥ 0.50 (≥ 0.60) for matches we found: 0 (one) missing intra-group matches for G1.1, two (three) missing intra-group matches for G1.2, one (two) missing intra-group matches between two samples of G1.3, zero (zero) missing intra-group matches for G2 and four (zero) missing intra-group matches for G3. G1.1 and G1.2 matched in 23 (22) over 25 pairwise calculations. G1.1/G1.2 had 21 (0) false positive matches with G3 and 0 false positive matches with the remaining samples. The remaining inter-group pairwise calculations had PCCs < 0.50. These samples and their correlation coefficients were used to create the PCoA multidimensional maps (Fig. 2) and used later as the active matrix. Similarly, results of the chi² cross-correlation are shown in Supp. Fig. 2 and its PCoA in Supp. Fig. 3.

Using the automatic selection approach, we found that 10 of all the samples found in three recordings had PCCs < 0.50 concerning all samples from the active matrix (Table 2) thus forming new sound types. Concerning these samples, six out of 10 had PCCs ranging from 0.52 to 0.83 as the absolute highest value reached for intra-group pairwise calculations in this analysis. These samples were included in a new sound type, i.e. "G4.2", while the other four, i.e. "G4.1", "G5.1", "G5.2" and "G5.3", were different to all other sound types including G4.2 (Table 1).

The PCCs with chosen samples from the Icelandic repertoire are also shown in Table 2. Of the nine samples, five had PCCs ≥ 0.50 and two ≥ 0.60 indicating low and high positive matches to our samples including manual and automatic approaches.

**Discussion**

In this analysis, we distinct a total of 10 sound types including nine calls and one composite sound as part of the visually and aurally created repertoire of the orca pod. We verified that the machine signal matching consisted of FFT cross-correlation using the Pearson methods and PCoA can successfully distinguish between different sound types. The efficiency of identification concerning the manual approach using the Pearson cross-correlation is > 85% (> 70%) for the 0.50 (0.60) threshold and the method appears to be less sensitive than expected in distinguishing similar sound types (e.g. G1.1 and G1.2). Results of the chi² cross-correlation (Supp. Fig. 2) successfully clustered sound types but the PCoA from the former method has better quality, i.e. points belonging to a sound type are closer to each other in space and the group of points belonging to the different types are more spaced, over the latter (Fig. 2; Supp. Fig. 3).

*Visual analysis*

Judging from the different occurrence of the sound types, it could be assumed that some of the call types are more frequent, e.g. G1.1 (16%) and G1.3 (24%), while others are rare, e.g. G7.2 (5%) and G8 (3%), which may indicate that some calls have a more general-purpose (group identification call) while others are more specific.

Concerning differences in certain sound types, e.g. G1.3 and G1.8, the emitter-to-hydrophone geometry has to be taken into account. As an example, composite sounds including clicks, which may have a directional part, might result in different sound types depending on the direction from which the sound was recorded. Therefore, the absence of clicks in G1.3 could be due to the hydrophone being off-axis from the animal. Alternatively, we observed the repeated sequences of the calls starting from a simplified version (G1.3) and then developed to reach a more complex structure (G1.8) as a possible communicational behaviour. The presence/absence and shape of HFCs in the samples is likely caused by individual-specific sound generation. For example, G1.1, G1.2 and G1.3 could be grouped into the same sound type produced by three over four individuals in the pod. However, without any visual confirmation, this can not be proved.

Concerning the patterns we found in the recordings, a call-answer in communication between two individuals is intriguing. Alternative explanations might involve a complicated structure similar to a phrase generated by a single individual. However, again in absence of visual confirmation, this remains speculative.

The selected samples from the Icelandic catalogue are generally similar to our sound types in terms of the general shape of the spectrogram, although the SBI is different as well as the trend of HFCs, which can be attributed to group-specific dialects.

Compared to our samples, those from the Antarctic catalogue are generally characterised by higher SBIs and clicks often forming a composite sound as the first part of it while, concerning our sounds, clicks generally follow calls. A roughly similar proportion of sound types were visually similar to those from our pod, compared to the same proportion with the Icelandic catalogue, which is in contrast with our expectations. Without performing a cross-correlation between samples, we cannot state whether there was actual (di)similarity using a more detailed machine comparison.

*Cross-correlation analysis*

Concerning the manual approach, we found only one false positive match, i.e. G1.1 and G1.2 vs G3, probably due to very dense LFCs with similar SBIs matching in some regions because of varying trends over time (Fig. 2).

The cross-correlation method with the automatic sample selection from the stream revealed six samples forming a new sound type (i.e. G4.2) and four more distinct sound samples (i.e. G4.1, G5.1, G5.2 and G5.3). The three samples of sound type G5 were formerly attributed to the same class of this signal in the visual and aural identification due to the similar LFC trends and roughly similar SBIs. The oscillating frequency patterns present in the samples are visually slightly different, which likely leads to lower PCCs. However, it is not clear if the signals are differently modulated on purpose or consist of the same call with natural variations.

When comparing our partial repertoire to that from Iceland, we found two high positive matches and five low positive matches. In particular, the match of I53.1 with G4.2 resulted for all six pairwise calculations with PCCs ≥ 0.60, which can be confirmed both visually and aurally, although the initial parts are slightly different (Supp. Fig. 4). This finding might be a confirmation that this pod is part of the Icelandic clan. However, it is not yet known if different clans may have such similarities, also considering the low complexity of this sound type. I53.1 also showed a high positive correlation to G4.1 while the latter and G4.2 were different (PCCs ranging from 0.19 to 0.45). I43.1, which was the only composite sound of the Icelandic catalogue, indicated a low and high positive correlation to G1.1/G1.2 and G3, respectively. The first part of I43.1 was visually and aurally different from the three sound types while the part consisting of increasing dense LFCs was visually similar to G1.1/G1.2 rather than G3, which had decreasing LFCs. I42.1 showed a low positive correlation to G1.1/G1.2 and G3 but the samples did not match visually and aurally suggesting the sound types were different. Therefore, this result could be considered as a false positive match. The low positive correlation between I53.2 and G1.3 was likely due to different HFCs, while LFCs and SBIs were similar according to the spectrograms. These sounds were also similar aurally. Finally, I71 indicated a low positive correlation to G3 as the SBI of these two samples partly matched because the former lasted more than the latter, which was generally true for all Icelandic samples.

*Study limitations*

In this study, we faced issues concerning computational power and data quality. As already mentioned, we just used 16 out of 217 five-minute recordings with orca sounds for the analysis, because they were quite representative of the entire data, i.e. they contained most of 33 sound types from the preliminary classification as well as a high number of samples for each type. When the inclusion of all recordings might provide more accurate results in terms of several sound types as well as their classification due to bigger samples. However, not all the recordings were prone to be analysed because of low quality, thus low SNR values potentially leading to biased matches. In conclusion, by keeping just samples with SNR $\geq 5$, most recordings would be automatically discarded resulting in a reduced number of those available for the sound type classification process. Regarding cross-correlations, the parameters we set potentially influenced the results. In particular, the bandpass filter of 0.2–9 kHz partly excluded both LFCs and HFCs in spectrograms of the detected samples (Fig. 1) thus limiting the area within which similarities were step-computed by the PCC cross-correlation. For example, lowering the upper limit of the bandpass filter to 5 kHz would probably lead to a high positive match for sound types such as I53.2 and G1.3 because the HFC would be discarded thus impeding detection of actual differences. However, a wider bandpass filter would rather include a larger area of background noise matrix, which is equal for all samples thus resulting in false positives concerning similarity, over spectrogram details. This was particularly true for our recordings with faint spectrograms after the peak of intensity occurring from 0 to 10 kHz.

**Conclusions**

The data continuously collected over several days for the orca pod occurring in the vicinity of the Pra's harbour represented a unique opportunity to study pod repertoire and test machine sound classification methods. Our survey was conducted following canonical visual and aural sound classification followed by cross-correlation methods. The Pearson cross-correlation method appears to be efficient (> 85%) with a low but non-negligible number of false positives using a threshold of 0.50.

We found a high positive match for one detected sound type (G4.2, 6 samples) with one from the Icelandic catalogue (I53.1), which was confirmed visually and aurally. The comparisons of various catalogues using cross-correlation methods appear to be a non-trivial task due to the different data acquisition instruments and environmental conditions. In particular, different sampling rates make cross-correlations hardly possible so we suggest 192 kHz as a sampling rate for the sound catalogues. In the used catalogues, each sound type had often only one sample, allowing limited possibilities to verify the match and distinguish false positives. More samples for each sound class are desired which is not an easy task for the catalogue creation using boat campaigns.

The background presence is a known limitation for the Pearson methods. Since this method technically consists of a linear fit, it is expected that a signal with monotonic frequency lines would match with a signal having frequency lines oscillating around the frequency lines of the former signal. This, however, seemed not to be a limitation when comparing sound types such as G5.1, G5.2 and G5.3. To improve the reliability of cross-correlation outputs, we tried to implement $chi^2$ for pairwise calculation. Our preliminary implementation is rather promising, but it performs worse concerning Pearson methods. One of the limitations of the method we used is that the background is not automatically subtracted, which is desired for future implementations (e.g. considering a nearby quiet sample). This holds also for the Pearson methods although this is less influenced by the background noise matrix.

Our first attempts to classify the orcas sounds using machine techniques are yet far from being a standalone procedure, limited by the computational complexity and arguable results including mismatches and false positives. Furthermore, it looks like orca vocalisations are way more complex, including differences in the repetition of one sound by different individuals or even by the same individuals, than bird vocalizations, for which warbleR was developed. However, we believe that cross-correlation methods already represent a powerful tool for sound classification when accompanied by aural and visual comparison. We expect more developments will come in this field leading to a human-free and unbiased data analysis for repertoire compilation and identification of species as well as groups and clans.

**Authors contribution**

This submission represents original work carried out by the authors. All authors have reviewed and approved the manuscript and agree to its submission. Conceptualisation: S.P., C.G., V.K., M.San., methodology: S.P., C.G., V.K., M.San., data collection: M.M., formal analysis and investigation: S.P., C.G., V.K., M.San., writing—original draft preparation: S.P., V.K., writing and editing: S.P., C.G., V.K., M.M., M.San.

**Acknowledgements**

The study was funded by the University of Genova. We greatly thank NAUTA scientific for providing instrumentation and sharing acoustic data and the Tethys Research Institute for the data collection. Special thanks to the Coast Guard and Port Authority for the administrative and technical support. We are also grateful to the staff of Acquario di Genova and Marineland (Antibes) for the fruitful discussions.

**Conflict of interests**

The authors declare that they have no known competing financial interests or personal relationships that could have appeared to influence the work reported in this paper.

**Data accessibility**

Further access to raw data is offered by NAUTA scientific, Italy, to other research groups willing to test their procedures and analysis methods over a known reference dataset.

**Figures and tables**

| Call group | Input number of samples | Number of selected samples (% of the input) | SNR >5 (% of the input) | Cluster size for intra-group cross-correlation |
|---|---|---|---|---|
| G1.1 | 25 | 22 (88%) | 9 (36%) | 5 |
| G1.2 | 15 | 14 (93%) | 11 (73%) | 5 |
| G1.3 | 37 | 34 (92%) | 22 (59%) | 5 |
| G2 | 27 | 19 (61%) | 4 (15%) | 4 |
| G3 | 30 | 26 (87%) | 12 (40%) | 5 |

**Table 1** Output of the auto_detec function selecting samples from the dataset including 16 recordings with five sound types.

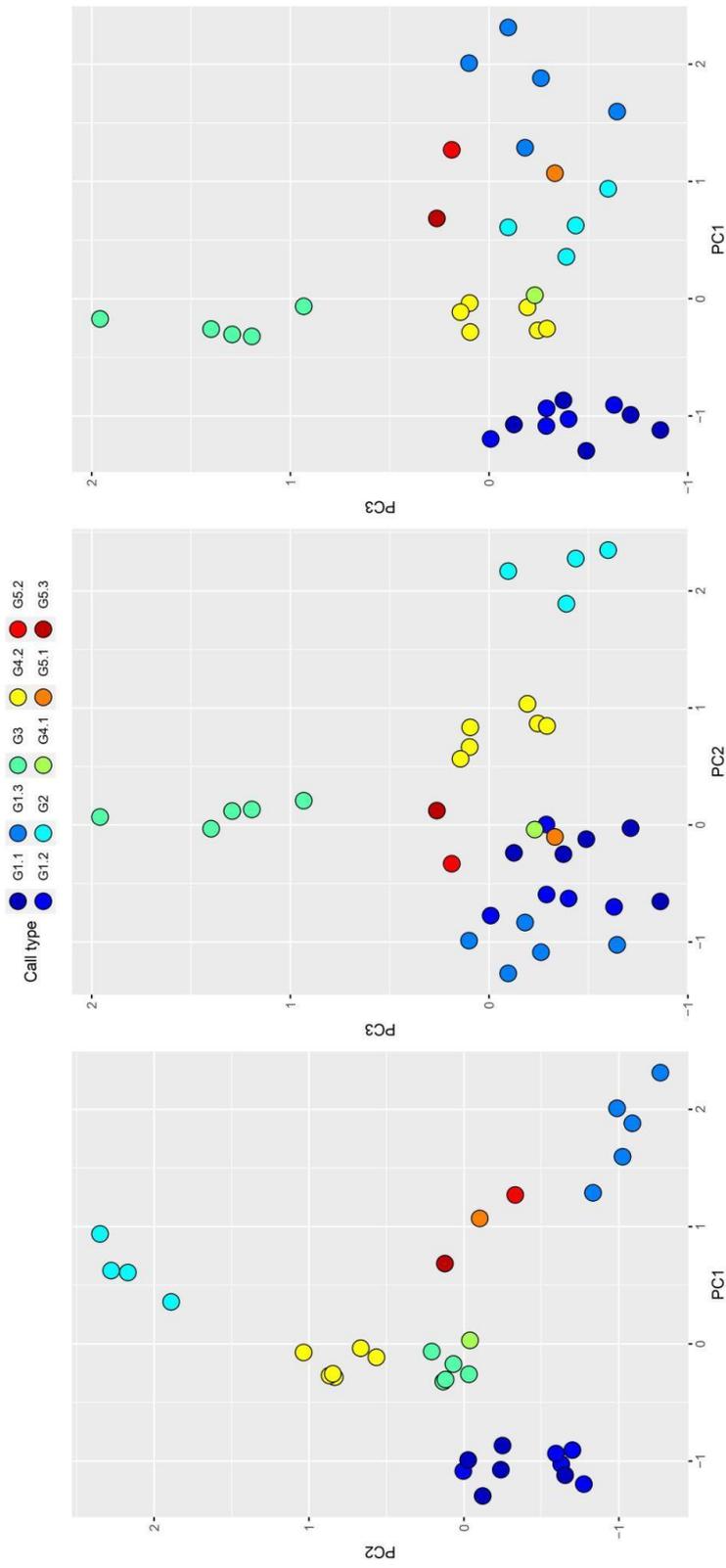

**Fig. 2** PCoA obtained from the Pearson cross-correlations for all sounds of the first and second approach

| Sound type (cluster size) | G1.1 | G1.2 | G1.3 | G2 | G3 | G4.2 | G4.1 | G5.1 | G5.2 | G5.3 |
|---|---|---|---|---|---|---|---|---|---|---|
| G1.1 (5) | **0.50–0.82** | 0.39–0.78 | 0.17–0.46 | 0.16–0.38 | 0.31–0.58 | 0.31–0.44 | 0.38–0.43 | 0.29–0.36 | 0.23–0.31 | 0.19–0.25 |
| G1.2 (5) | 0.39–0.78 | **0.44–0.72** | 0.14–0.42 | 0.21–0.43 | 0.28–0.57 | 0.24–0.43 | 0.35–0.42 | 0.24–0.34 | 0.18–0.29 | 0.10–0.27 |
| G1.3 (5) | 0.17–0.46 | 0.14–0.42 | **0.48–0.82** | 0.14–0.35 | 0.23–0.45 | 0.12–0.28 | 0.26–0.38 | 0.31–0.49 | 0.34–0.47 | 0.20–0.28 |
| G2 (4) | 0.16–0.38 | 0.21–0.43 | 0.14–0.35 | **0.60–0.67** | 0.22–0.43 | 0.22–0.44 | 0.24–0.32 | 0.23–0.32 | 0.17–0.26 | 0.16–0.25 |
| G3 (5) | 0.31–0.58 | 0.28–0.57 | 0.23–0.45 | 0.22–0.43 | **0.56–0.77** | 0.34–0.38 | 0.25–0.36 | 0.25–0.36 | 0.28–0.34 | 0.22–0.27 |
| G4.2 (6) | 0.31–0.44 | 0.24–0.43 | 0.12–0.28 | 0.22–0.44 | 0.22–0.43 | **0.52–0.83** | 0.19–0.45 | 0.22–0.27 | 0.17–0.26 | 0.15–0.37 |
| G4.1 (1) | 0.38–0.43 | 0.35–0.42 | 0.26–0.38 | 0.24–0.32 | 0.34–0.38 | 0.19–0.45 | — | 0.25 | 0.26 | 0.49 |
| G5.1 (1) | 0.29–0.36 | 0.24–0.34 | 0.31–0.49 | 0.23–0.32 | 0.25–0.36 | 0.22–0.27 | 0.25 | — | 0.48 | 0.20 |
| G5.2 (1) | 0.23–0.31 | 0.18–0.29 | 0.34–0.47 | 0.17–0.26 | 0.28–0.34 | 0.17–0.26 | 0.26 | 0.48 | — | 0.14 |
| G5.3 (1) | 0.19–0.25 | 0.10–0.27 | 0.20–0.28 | 0.16–0.25 | 0.22–0.27 | 0.15–0.37 | 0.49 | 0.20 | 0.14 | — |
| I40.1 (1) | 0.22–0.37 | 0.29–0.37 | 0.15–0.25 | 0.05–0.12 | 0.13–0.23 | 0.06–0.15 | 0.11 | 0.09 | 0.05 | 0.00 |
| I42.1 (1) | 0.36–0.55 | 0.32–0.50 | 0.31–0.39 | 0.32–0.40 | 0.48–0.59 | 0.36–0.48 | 0.39 | 0.24 | 0.28 | 0.20 |
| I43.1 (1) | 0.24–0.58 | 0.27–0.50 | 0.09–0.26 | 0.34–0.44 | 0.39–0.61 | 0.31–0.45 | 0.41 | 0.34 | 0.24 | 0.18 |
| I47 (1) | 0.24–0.38 | 0.18–0.27 | 0.21–0.39 | 0.16–0.24 | 0.26–0.37 | 0.18–0.26 | 0.37 | 0.31 | 0.29 | 0.35 |
| I53.1 (1) | 0.37–0.47 | 0.34–0.45 | 0.21–0.29 | 0.34–0.44 | 0.40–0.46 | 0.62–0.67 | 0.67 | 0.24 | 0.24 | 0.25 |
| I53.2 (1) | 0.25–0.36 | 0.23–0.39 | 0.46–0.55 | 0.21–0.23 | 0.26–0.32 | 0.18–0.30 | 0.40 | 0.32 | 0.40 | 0.22 |
| I62.1 (1) | 0.14–0.27 | 0.17–0.31 | 0.05–0.20 | 0.14–0.21 | 0.10–0.20 | 0.09–0.12 | 0.18 | 0.36 | 0.19 | 0.11 |
| I64 (1) | 0.34–0.42 | 0.28–0.40 | 0.29–0.38 | 0.28–0.37 | 0.42–0.48 | 0.29–0.41 | 0.40 | 0.31 | 0.27 | 0.19 |
| I71 (1) | 0.32–0.41 | 0.31–0.37 | 0.14–0.22 | 0.26–0.28 | 0.25–0.38 | 0.21–0.28 | 0.20 | 0.18 | 0.16 | 0.06 |

**Table 2** Summary of the intra- and inter-group PCCs ranges resulting from the cross-correlation for various signal types in the studied pod as well as those from the Icelandic catalogue. Inter-group PCCs exclude self-values which would be 1. Pairwise calculations between the Icelandic samples are not shown in the table.

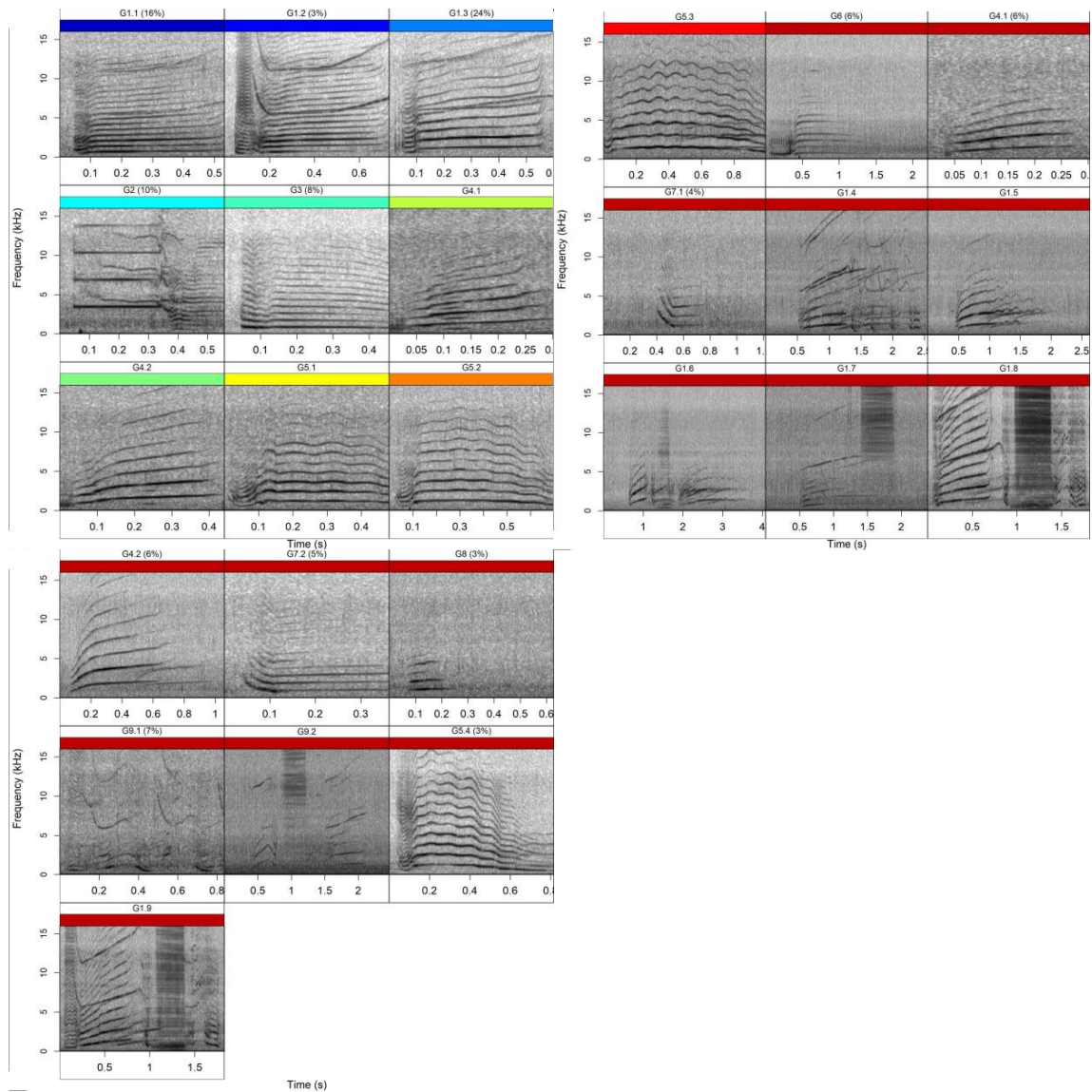

**Fig. 1** Spectrograms of all 25 sound types from our pod identified in this work including preliminary analysis, manual and automatic approach. The percentage between brackets indicates sound type occurrence and it is not shown concerning composite sounds and sound types resulting from the automatic approach. In the first case, the percentage value of the sound types forming a composite sound includes that of the latter. Sound types in dark red refer to the preliminary analysis. G4.1 and G4.2 are shown two times because they were both found in the preliminary analysis (with %) and with the automatic approach. The greyscale shows the signal amplitude (-80 dB - 0 dB) versus time (x-axis) and frequency (y-axis). The background noise contribution is seen around 0 kHz. The noise variation seen as overall spectrogram background change between different signals is due to the environmental noise variation with time.

**Supplementary Information**

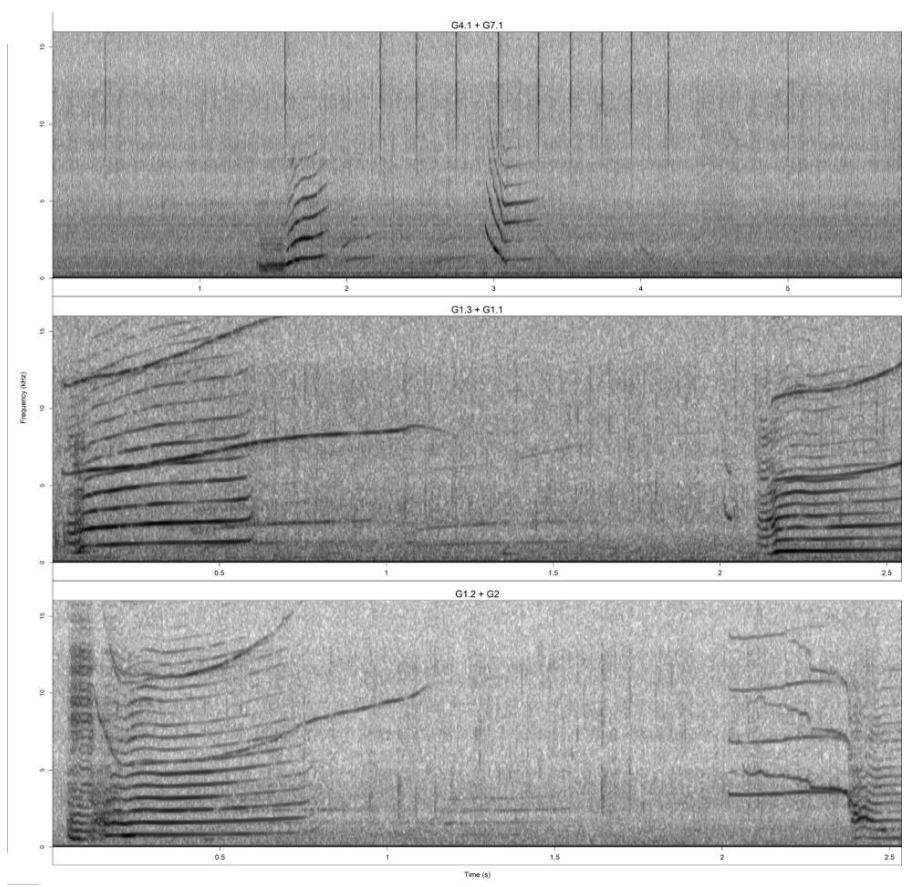

**Supp. Fig. 1** Spectrograms of the three patterns found in the analysed recordings. The first image showing G4.1 followed by G7.1 was also found with G4.2 followed by G7.2.

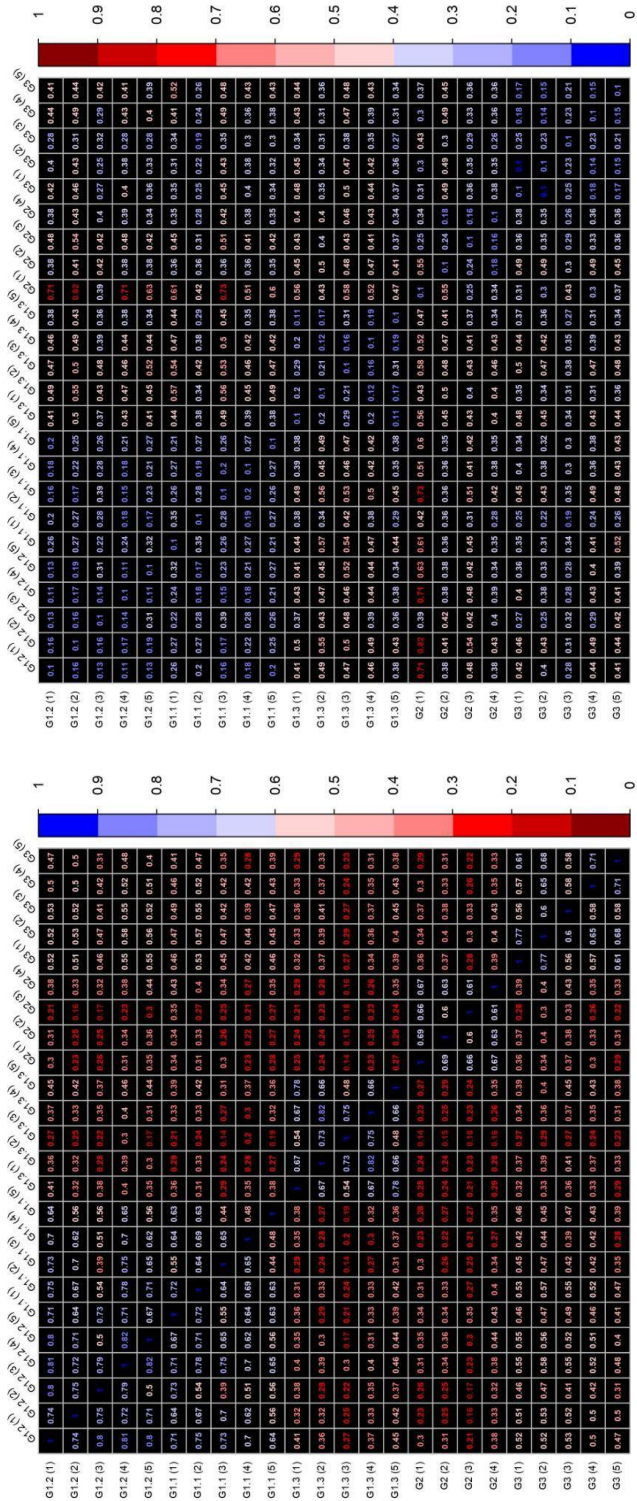

**Supp. Fig. 2** Correlation coefficients resulting from pairwise Pearson (left) and chi² (right) calculations using the five sound types. On the right table, chi²/10 is plotted for visual purposes in the R package.

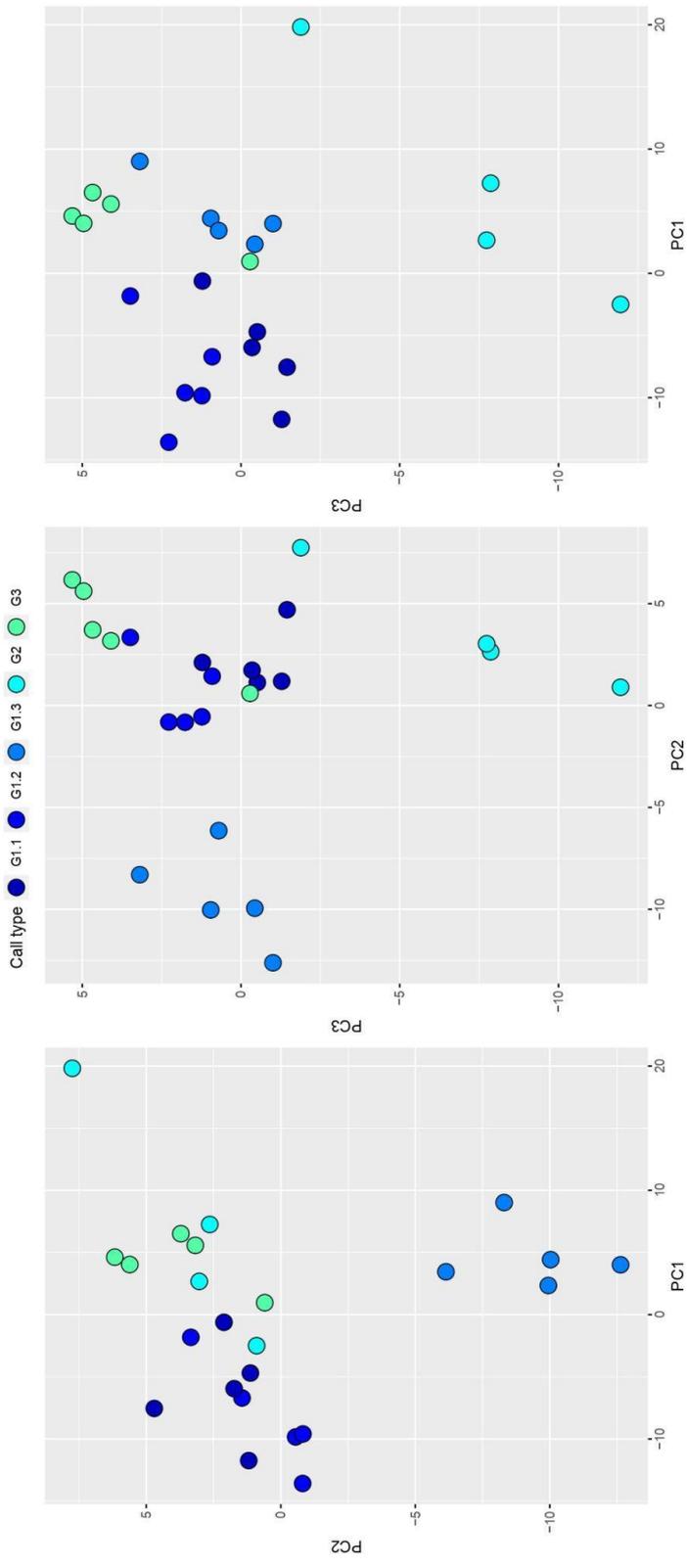

**Supp. Fig. 3** PCoA obtained from the chi² cross-correlation of the five sounds of the first approach

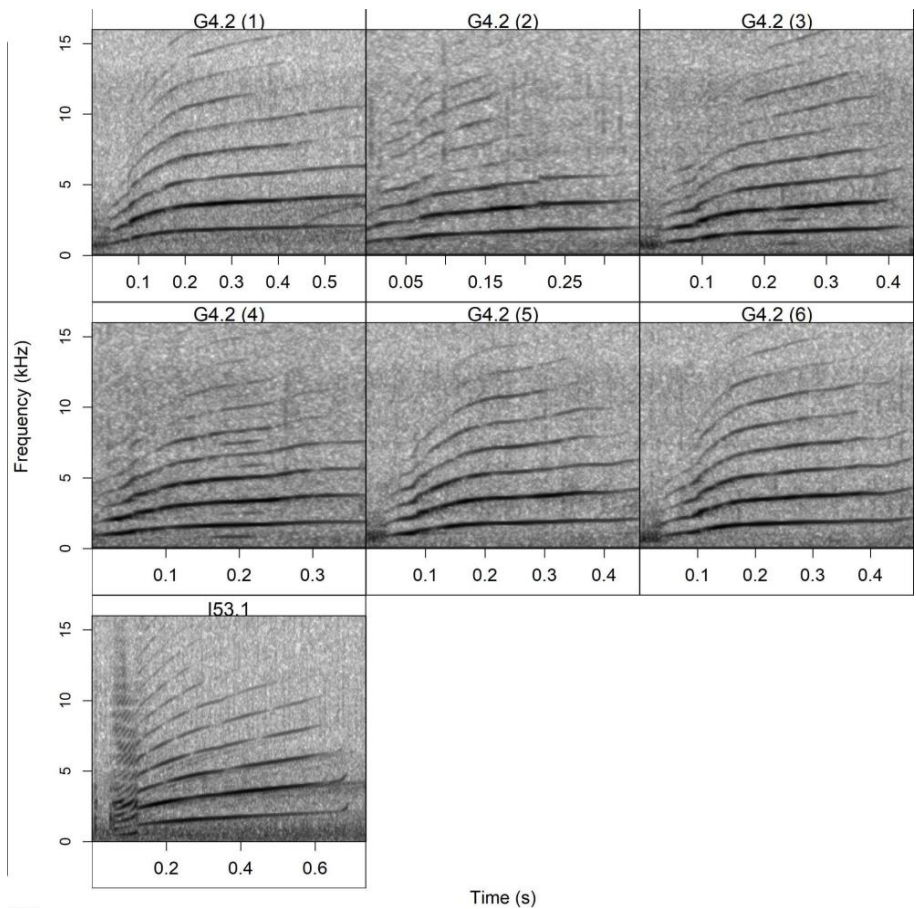

**Supp. Fig. 4** Spectrograms of the six samples belonging to sound type G4.2 and I53.1. Among all PCC resulting from the comparison between the repertoire from this study and that from Iceland, these two sound types had the highest absolute values. The greyscale shows the signal amplitude (-80 dB - 0 dB) versus time (x-axis) and frequency (y-axis). The background noise contribution is seen around 0 kHz. The noise variation seen as overall spectrogram background change between different signals is due to the environmental noise variation with time.